\documentclass[conference]{IEEEtran}

\IEEEoverridecommandlockouts
\usepackage{cite}
\usepackage{amsmath,amssymb,amsfonts}
\usepackage{algorithmic}
\usepackage{graphicx}
\usepackage{textcomp}
\usepackage{xcolor}
\usepackage{float}
\usepackage{multirow}
\usepackage[hidelinks]{hyperref}
\def\BibTeX{{\rm B\kern-.05em{\sc i\kern-.025em b}\kern-.08em
    T\kern-.1667em\lower.7ex\hbox{E}\kern-.125emX}}

\begin{document}
\title{Impact of Work Schedule Flexibility on EV Hosting Capacity: Insights from Analyzing Field Data}
\author{
    \IEEEauthorblockN{Marco Iorio, Mohammad Golgol, and Anamitra Pal}
    \IEEEauthorblockA{\text{School of Electrical, Computer, and Energy Engineering}, \textit{Arizona State University}, {Tempe, AZ}\\
    miorio2@asu.edu, mgolgol@asu.edu, anamitra.pal@asu.edu}
}

\maketitle

\begin{abstract}
Uncoordinated electric vehicle (EV) charging is altering residential load patterns and pushing distribution transformers to operate beyond their limits.
These outcomes can be offset by 
exploiting 
the flexibility in work schedules (hybrid, remote vs. in-person) of EV owners, particularly when combined with rooftop
photovoltaic (PV) generation. 
However, this phenomenon
has not been explored in-depth yet. 
This paper addresses this research gap by introducing
weekly work schedule-aware robust and chance-constrained optimization formulations for EV charging coordination
to determine a
transformer's EV hosting capacity.
The results obtained using data from a
residential feeder in Arizona indicate that 
an intelligent combination of work schedule flexibility with PV generation can help power utilities effectively manage changing grid demands.
\end{abstract}


\section{Introduction}\label{intro} 


In the era of transportation electrification, the power grid is responsible for supplying the charging power to exceedingly more high-capacity electric vehicles (EVs). However, in EV-dense neighborhoods, charging during peak demand periods can overload distribution transformers and intensify regional grid stress \cite{roy2023evimpact}. Given 80\% of charging is done at home in the United States, EV integration solutions must be (i) coordinated to minimize burden on grid assets,
and (ii) convenient enough for EV owners to adopt \cite{Home_Charging}. 
This paper explores the combination of \textit{work schedule-aware charging solutions} with \textit{time-of-use (TOU) price plans} and \textit{on-site rooftop solar generation} to determine EV hosting capacity (HC) of transformers as well as identify transformers in need of upgrades.

The proposed framework explicitly models three distinct work arrangements: \textit{in-person}, \textit{hybrid}, and \textit{remote}, each formulated in \textit{robust} and \textit{chance-constrained} forms to capture behavioral variability and uncertainty. Leveraging field data from residential transformers with varying solar photovoltaic (PV) penetration, this study assesses how work-schedule diversity impacts a transformer's EV HC to
inform the design of new demand-flexibility programs for power utilities. 

Another unique feature of this study is the coordination of EV home charging over the period of a week.
Prior studies have often focused on daily charging horizons, assuming that each EV must satisfy its full energy demand within a single day \cite{jointmodelling_ev_charging,al-alwash2024optimization,Nick_OAJPE_Paper}.
In practice, however, most EVs deplete their batteries over several days. The growing diversity in modern-day work arrangements and charging patterns 
suggests that satisfying charging requirements across a weekly horizon offers a more realistic and flexible strategy. 
In summary, the central contributions of this paper are as follows:
\begin{itemize}
    \item Introduce a weekly work schedule-aware EV charging optimization framework that models in-person, hybrid, remote, and mixed (which comprises all three) work arrangements, revealing how mobility patterns shape untapped demand-flexibility and solar-alignment potential.

    \item Develop a novel robust and chance-constrained formulation for estimating EV HC, enabling a direct comparison of conservative and probabilistic planning strategies using real-world transformer data from Arizona.
    
    \item Demonstrate how coordinated EV charging can mitigate mid-day PV-driven reverse power flows, quantify the additional HC unlocked by rooftop-PV penetration, and identify transformers most sensitive to work-arrangement demographic composition and load variability.
\end{itemize}
\section{Model Development and Considerations}
\label{method}

The in-person, hybrid, and remote priority-weighted EV charging optimization models are designed to provide hierarchical priority to charging at certain days/times of the week.
Together, the three models present a new suite of weekly work schedule-aware EV-TOU price plans that utilities can package to customers.
The details of the hierarchical charging priority associated with each work schedule-aware EV charging optimization model is presented below.



For the in-person model, weekdays during super off-peak or weekday night hours are assigned the highest weight and lowest cost, followed by weekends during typical PV generation hours, and then weekends outside of PV generation hours. This framework accommodates EV owners who are unable to charge their car from home during typical solar generation periods throughout the work week.

For the hybrid model, weekends during typical PV generation hours are assigned the highest weight and lowest cost, followed by designated work-from-home (WFH) weekdays during PV generation hours, and then weekdays during super off-peak hours. This structure accommodates EV owners who are available to charge their car from home on 
WFH weekdays.

For the remote model, weekdays during typical PV generation hours are assigned the highest weight and lowest cost, followed by weekends during PV generation hours, and then weekends outside of PV generation hours. This framework accommodates EV owners who are available to charge their car from home any day of the work week.

\vspace{-0.5em}
\subsection{Optimization Model and Data Preliminaries}

The analysis was conducted using data from an SRP feeder in Arizona for the month of July.
As such, SRP’s summer peak EV-TOU price plan \cite{srp_ev_price_plans} was expanded over a week and adopted into the coordination models.
The weight-based constraints help each model
select charging times that have the highest priority.
Therefore, the higher weighted times throughout the week were inversely
proportional to the cost of electricity.

The rated charging power was assumed to be 7.2 kW, the typical power delivered by most 240-volt AC residential electric vehicle supply equipment (EVSE) \cite{AFDC_EVchargingStations}. The charging efficiency was assumed to be 80\%, which includes the efficiency of (i) the charging process from the home-based EVSE, (ii) the EV's onboard AC-DC charger, and (iii) charging the lithium-ion battery cells \cite{EPA_FuelEconomyEVRangeTesting_2025}.
The average energy consumption was assumed to be 0.32 kWh/mile for all EVs \cite{EPAWhatIfOneOfYourCarsWasElectric}.

\vspace{-0.5em}
\subsection{Parametric Uncertainty Quantification}
\label{ParamUncerQuan}


All EVs were considered to be fully electric
in this study. The \textit{battery capacity} of each EV was assumed to follow a chi-squared right-skewed distribution ranging from 77 to 118 kWh. This is justified when ranking the top selling EVs of November 2024 by their popularity and usable battery capacity \cite{driveelectric2025evminute, AFDC_ModelYear2024,EVGuide_BatteryCapacityEstimatingRange}. 
The initial 
\textit{state-of-charge} (SoC) at the start of the week for each EV is assumed to follow a chi-squared right-skewed distribution ranging from 80\% to 95\%. This comes from the data-driven recommendation to charge EVs up until 80\% SoC to reduce charging losses and maintain battery longevity \cite{KOSTOPOULOS2020}.
The initial SoC is set
to be equal to the final SoC to devise a continuous representation of EV charging scheme for every
week. 
The daily \textit{one-way driving distance} for each morning and evening commute is assumed to be a chi-squared left-skewed distribution ranging from 27 to 37 miles. This is in accordance with the 2022 National Household Travel Survey's average vehicle trip length on weekdays \cite{NHTS_2022}.

For the \textit{robust} formulation, the EV battery capacity, initial/final SoC, and one-way driving distance are assigned the most extreme values of their respective distributions. 
The constraints are also made to satisfy with a 100\% guarantee. 
This formulation corresponds to the worst-case design and forms the baseline for comparison purposes. 
The \textit{chance-constrained} formulation presents
a 
stochastic design where each uncertain parameter is driven by a scenario-based Monte Carlo simulation. The constraints associated with the stochastic parameters are
satisfied with a high probability, but with $<$100\% guarantee. This formulation 
captures
the realistic behavioral uncertainties that underlie EV charging patterns.

\section{Mathematical Formulation}
\label{math form}

The mathematical formulations of the robust and chance-constrained problems aim to coordinate EV home charging over a week to minimize charging expenses while ensuring no transformers are overloaded.
The work arrangement models share the same generalized setup with key differences being weight-based constraints, charging availability periods, driving periods, and EV-TOU cost periods throughout the week.





The objective function \eqref{eq:1} minimizes the total weekly charging cost for all EVs connected to a specific transformer with higher weighted days given more 
priority:
{\normalsize
\begin{equation}
    \min \sum_{v=0}^{|\mathcal{V}|-1} \bigg( \sum_{t=0}^{|\mathcal{T}|-1} u \cdot \eta \cdot b[v,t] \cdot \frac{1}{w[t]} \cdot c[t] \bigg)
    \label{eq:1}
\end{equation}
}

In \eqref{eq:1}, $\mathcal{V}$ is the set of all EVs, $\mathcal{T}$ is the set of all modeled weekly time periods (Mon-Sun) each with 15-minute intervals ($\mathcal{T}=\{0,1,\ldots,\mathcal{T}_{\max}\}$, where $\mathcal{T}_{\max}$ is the last time period of
the week), $u$ is the rated charging power in kW, $\eta$ is the charging efficiency, $w[t]$ is the weight parameter assigned to each time period, and $c[t]$ is the EV-TOU cost parameter assigned to each time period. The binary variable $b[v,t]$ has a value of $1$
if an EV is charging at time period $t$. 

Constraints \eqref{eq:weights-pt-1} and \eqref{eq:weights-pt-2} correspond to the in-person charging priority order (see Section \ref{method}).
These two constraints ensure that there are more charging instances for the higher priority-weighted hours of the week compared to the lower priority-weighted hours of the week. The term $w[t]$ was scaled by a factor of 10 between each priority order subset in the week to better distinguish between these specific charging intervals.
{
\begin{IEEEeqnarray}{rCl}
\begin{aligned}
\sum_{\forall t \in \{\text{Weekday Nights}\}} b[v,t] \cdot w[t]  & \geq \sum_{\forall t \in \{\text{Weekend PV}\}} b[v,t] \cdot w[t]
\quad \\ & \forall v \in \mathcal{V}
\end{aligned}
\label{eq:weights-pt-1}
\end{IEEEeqnarray}
}
\vspace{-1em}
{
\begin{IEEEeqnarray}{rCl}
\begin{aligned}
\sum_{\forall t \in \{\text{Weekend PV}\}} b[v,t] \cdot w[t] & \geq \sum_{\forall t \in \{\text{Weekend No PV}\}} b[v,t] \cdot w[t]
\quad \\ & \forall v \in \mathcal{V}
 \end{aligned}
 \label{eq:weights-pt-2}
\end{IEEEeqnarray}
}




Constraint \eqref{eq:charging-demand} establishes the charging demand required for each EV over the entire week for the \textit{robust} formulation: 
\begin{equation}
SoC^{i}[v]\; + \sum_{t=0}^{|\mathcal{T}|-1} \bigg( u \cdot \eta \; \cdot \;  b[v,t] - d[v,t] \cdot \delta \bigg) \geq SoC^{f}[v] 
\label{eq:charging-demand}
\end{equation}
\vspace{-1em}
\begin{equation*}
\forall v \in V 
\end{equation*}

In this expression, the binary variable $d[v,t]$ has a value of $1$
if the EV is driving at time period 
$t$. The energy depleted from driving, $\delta$, is computed by multiplying the average EV energy consumption per mile by the one-way commuting distance (in miles). 
The terms $SoC^{i}[v]$ and $SoC^{f}[v]$ represent the initial and final SoC, respectively, for each EV.

Constraint \eqref{eq:charging-demand} ensures that the total weekly charging process minus the energy depleted from driving satisfies the final SoC for each EV. 
However, this constraint must be reformatted in the \textit{chance-constrained} formulation
to handle the stochasticity in initial and final SoC and energy depleted from driving. Note that the stochasticity in EV battery capacity is inherently captured in the initial and final SoC random variables. 
To ensure a high guarantee of constraint satisfaction, an $\epsilon$ term is introduced to modify \eqref{eq:charging-demand} to get
\eqref{eq:stochastic-notation}:
\begin{IEEEeqnarray}{rCl}
& \mathbb{P} \Bigg\{SoC^{i}[v]\; + & \sum_{t=0}^{|\mathcal{T}|-1} \bigg (u \cdot \eta \cdot b[v,t] - d[v,t] \cdot \delta \bigg) \geq SoC^{f}[v] \nonumber
\\
&  & \forall v \in V \Bigg\} \geq (1-\epsilon)
\label{eq:stochastic-notation}
\end{IEEEeqnarray}

Constraint \eqref{eq:stochastic-notation} is then reformulated to consider the nature of realized scenarios within an uncertainty set driven by known probability distributions. This becomes expression \eqref{eq:stochatic-reformulation} where $\xi$ is a realized scenario in the uncertainty set $\Xi$. The big M method is introduced along with the binary variable $z[\xi]$ to control the status if a constraint is satisfied for each scenario.
{\normalsize
\begin{equation}
\begin{split}
& SoC^{i}[v,\xi] \; +  \sum_{t=0}^{|\mathcal{T}|-1} \bigg(u \cdot \eta \cdot b[v,t] - d[v,t] \cdot \delta[\xi] \bigg) \geq \quad \\ & SoC^{f}[v, \xi]  - M \cdot (1 - z[\xi])  \quad  \forall v \in V, \; \forall \xi \in \Xi
\end{split}
\label{eq:stochatic-reformulation}
\end{equation}
}

For notational simplicity, \eqref{eq:SoC-rule-progress} introduces a new variable called $SoC^{\mathrm{rule}}[v,t,\xi]$ that recursively tracks the SoC progression for each EV at each time period under each uncertainty scenario:  
{\normalsize
\begin{equation}
\begin{split}
& SoC^{{\mathrm{rule}}}[v,t,\xi] = SoC^{{\mathrm{rule}}}[v,t-1,\xi]+ u \cdot \eta \cdot b[v,t] - \\ & \quad \quad \quad \quad \quad \quad d[v,t] \cdot \delta[\xi]   \quad  \forall v \in V, \; \forall t \in \mathcal{T}^{+},\; \forall \xi \in \Xi 
\end{split}
\label{eq:SoC-rule-progress}
\end{equation}
}

In \eqref{eq:SoC-rule-progress}, $\mathcal{T}^{+}$ is a subset of $\mathcal{T}$ where $\mathcal{T}^+=\{1,2,\ldots,\mathcal{T}_{\max}\}$. 
Constraint \eqref{eq:SoC-20percent-rule} inspired from \cite{KOSTOPOULOS2020} ensures that the SoC progression of each EV throughout the week does not fall below the 20\% threshold, $\alpha$, of its maximum usable energy capacity $B^{\max}[v,\xi]$.
This constraint is also 
stochastic
because it factors in the usable battery capacity of each EV (see Section \ref{ParamUncerQuan}). 
{\normalsize
\begin{equation}
\begin{split}
& SoC^{{\mathrm{rule}}}[v,t,\xi] \geq \alpha \cdot B^{\max}[v, \xi] - M \cdot (1-z[\xi]) \\ & \quad  \quad \quad \quad \quad \quad \forall v \in V, \; \forall t \in \mathcal{T}^{+},\; \forall \xi \in \Xi 
\end{split}
\label{eq:SoC-20percent-rule}
\end{equation}
}

Constraint \eqref{eq:SoC-exceed-rule} ensures the weekly SoC progression of each EV does not exceed its usable battery capacity:  
{\normalsize
\begin{equation}
\begin{split}
& SoC^{{\mathrm{rule}}}[v,t,\xi] \leq B^{\max}[v,\xi] - M \cdot (1-z[\xi]) \\ & \quad \quad \quad \quad \quad \quad \forall v \in V, \;  \forall t \in \mathcal{T}^{+},\; \forall \xi \in \Xi 
\end{split}
\label{eq:SoC-exceed-rule}
\end{equation}
}

Constraint \eqref{eq:constraint-guarantee} ensures that the summation of all instances of the variable $z[\xi]$ is greater than or equal to the $\epsilon$ constraint satisfaction guarantee for all the realized scenarios in the uncertainty set $\Xi$: 
{\normalsize
\begin{equation}
\begin{split}
& \sum_{\xi=0}^{|{\Xi}|-1} z[\xi] \geq (1 - \epsilon) \cdot |\Xi|
\end{split}
\label{eq:constraint-guarantee}
\end{equation}
}

Constraint \eqref{eq:trans-cap-limit} ensures that the combined home-based EVSE and household demand throughout the week does not exceed the transformer's rated power capacity
$S^{\max}$. 
In \eqref{eq:trans-cap-limit}, $H[t]$ is the aggregated household load at time $t$.
{\normalsize
\begin{equation}
\sum_{v=0}^{|\mathcal{V}|-1} \bigg(u  \cdot b[v,t]  + H[t] \bigg)\leq S^{\max} \quad \forall t \in \mathcal{T}
\label{eq:trans-cap-limit}
\end{equation}
}



Constraint \eqref{eq:min-charging-requirement} ensures that each instance of charging lasts for more than $\mathcal{T}^{\text{c}}_{\text{req}}$, where $\mathcal{T}^{\text{c}}_{\text{req}}$ is the minimum number of consecutive time periods each EV must remain in charging mode.
This prevents accelerated aging of the EV battery.
{\normalsize
\begin{equation}
\begin{split}
& b[v,t] - b[v,t-1] \leq b[v, t+i]  \quad \\ & \forall v \in \mathcal{V}, \; i \in [0, \mathcal{T}^{\text{c}}_{\text{req}}], \; t \in [1, \mathcal{T}_{\max} - \mathcal{T}^{\text{c}}_{\text{req}}]
\end{split}
\label{eq:min-charging-requirement}
\end{equation}
}

Constraint \eqref{eq:new-charge-session} prevents switching or initiating a new charging session if the remaining time periods
are less than $\mathcal{T}^{\text{c}}_{\text{req}}$. This also helps preserve the EV battery lifespan by limiting how many new charging sessions are allowed.
{\normalsize
\begin{equation}
\begin{split}
& b[v,t] \geq  b[v,t+1] \\ & 
\forall v \in \mathcal{V}, \; t \in \{ \mathcal{T}: \mathcal{T} \geq (\mathcal{T}_{\max} - \mathcal{T}^{\text{c}}_{\text{req}})\}
\end{split}
\label{eq:new-charge-session}
\end{equation}
}

Constraint \eqref{eq:min-driving-requirement} ensures that each instance of driving lasts for more than $\mathcal{T}^{\text{d}}_{\text{req}}$, where $\mathcal{T}^{\text{d}}_{\text{req}}$ is the minimum number of consecutive time periods each EV must remain in driving mode. In the simulations, both $\mathcal{T}^{\text{c}}_{\text{req}}$ and $\mathcal{T}^{\text{d}}_{\text{req}}$ were set to one hour (= four time periods).
{\normalsize
\begin{equation}
\begin{split}
& d[v,t] - d[v,t-1] \leq d[v, t+i]  \\ &
\forall v \in \mathcal{V}, \; i \in [0, \mathcal{T}^{\text{d}}_{\text{req}}], \; t \in [1, \mathcal{T}_{\max} - \mathcal{T}^{\text{d}}_{\text{req}}]
\end{split}
\label{eq:min-driving-requirement}
\end{equation}
}

Constraint \eqref{eq:mon-driving-limit} specifies the total duration of daily driving during the work week, which is limited to being below the variable,
$N$. 
The value of $N$ was set to
eight to simulate a morning and evening commute time of one hour each.
{\normalsize
\begin{equation}
\begin{split}
\sum_{t=0}^{|{\mathcal{T}_{i}}|-1} d[v,t] \leq N \quad 
\forall v \in \mathcal{V}, \; \forall i \in \{\text{Mon,...,Fri}\}
\end{split}
\label{eq:mon-driving-limit}
\end{equation}
}






Constraint \eqref{track-daily-new-charge} recursively tracks the number of changes in the charging status, with the variable $f[v,t]$ tracking the number of new daily changes: 
\begin{equation}
    f[v,t] \geq b[v,t] - b[v,t-1] \quad \forall v \in \mathcal{V}, \; \forall t \in \mathcal{T}^{+}
\label{track-daily-new-charge}
\end{equation}

Constraint \eqref{mon-charge-limit} limits the total number of daily charging switches to 
$Q$. 
The value of $Q$ was set to
three in the simulations to prevent premature degradation of the EV battery.
{\normalsize
\begin{equation}
\sum_{t=0}^{|\mathcal{T}_{{j}}|-1} f[v,t] \leq Q \quad \forall v \in \mathcal{V}, \; \forall j \in \{\text{Mon,...,Sun}\}
\label{mon-charge-limit}
\end{equation}
}

To summarize, \eqref{eq:1} and \eqref{eq:trans-cap-limit}--\eqref{mon-charge-limit} are shared among the robust as well as the chance-constrained formulations. Constraints \eqref{eq:weights-pt-1} and \eqref{eq:weights-pt-2} are modified based on the
work schedule-aware model.
Constraint \eqref{eq:charging-demand} and \eqref{eq:stochastic-notation} are exclusive to the robust formulation and chance-constrained formulation, respectively.
When the uncertainty set is considered, \eqref{eq:stochatic-reformulation}--\eqref{eq:constraint-guarantee} cater to the chance-constrained formulation. The same constraints catered to the robust formulation when the uncertainty set was dropped from consideration.


\begin{figure*}[ht]
  \centering
  \includegraphics[width=0.98\textwidth]{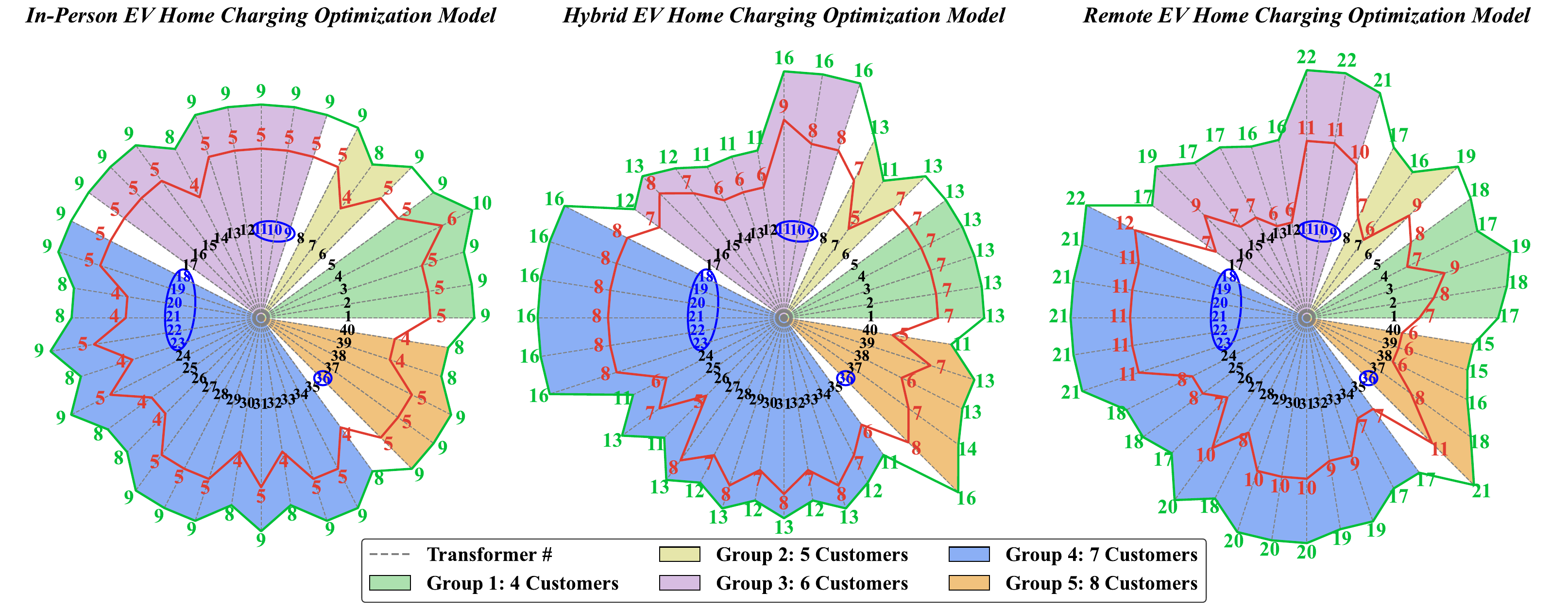}
   \vspace{-1em}
  \caption{EV HC results for robust and chance-constrained
  formulations of the in-person, hybrid, and remote EV charging optimization 
  models on 40 transformers at 50 kVA capacity each, located on an SRP feeder in Arizona. The numbers in black are transformers. The numbers in red are EV HC results for the robust formulation. The numbers in green are EV HC results for the chance-constrained formulation. Transformers circled 
  in blue ovals have loads with PV.}
  \label{fig:3_Hosting_Cap_Graphs}
  \vspace{-1em}
\end{figure*}

\section{Simulation Results}\label{results}

The proposed weekly charging coordination solutions were evaluated using field data sourced from SRP's 40 residential transformers, each of 50 kVA capacity. 
The goal is to determine the number of EVs
each transformer can reliably serve without getting overloaded under different work schedules.
The hybrid work schedule was implemented with one WFH day per week, while $\epsilon$ was $0.05$ and $|\Xi|$ was $1000$.
The EV HC results are presented as radial line graphs in Fig. \ref{fig:3_Hosting_Cap_Graphs}. Each dotted gray line represents a transformer, labeled $1$-$40$. The results for the robust and chance-constrained formulations are denoted by two concentric circles in red and green colors, respectively. Transformers supporting the same number of customers are grouped together with the same tint color (Groups $1$-$5$). The transformers supporting loads with PV are highlighted in blue.



\vspace{-0.25em}
\subsection{Transformer EV HC Under Different Work Schedules}
\vspace{-0.25em}

For the in-person EV charging optimization model (left plot of Fig. \ref{fig:3_Hosting_Cap_Graphs}), the EV HC ranges were 4-6 and 8-10 for the robust and chance-constrained formulations, respectively. This model was limited by the absence of charging flexibility during PV generation hours on weekdays seen in priority-weighted constraints \eqref{eq:weights-pt-1} and \eqref{eq:weights-pt-2}. Even with transformers supporting loads with PV, the HC range is noticeably small because solar-coordinated charging could only be leveraged on weekends. 

The hybrid EV charging optimization model (center plot of Fig. \ref{fig:3_Hosting_Cap_Graphs}) had EV HC ranges of 5-9 and 11-16, respectively, for the two formulations. This model, even with just one WFH weekday, yielded a significant improvement ($\approx$$60\%$) in the EV HC potential over the in-person model, particularly for those transformers that support loads with PV. This is because the temporal flexibility in one WFH day influences the model to shift charging to daytime PV generation hours.

For the remote EV charging optimization model (right plot of Fig. \ref{fig:3_Hosting_Cap_Graphs}), the EV HC ranges was 6-12 and 15-22 EVs, respectively; an increase of about 40\% over hybrid and 100\% over in-person models. The EV HC results for this model highlighted the positive effects of having charging flexibility during weekdays and weekends on accommodating 
more 
EVs.


\vspace{-0.5em}
\subsection{Effect of PV Penetration and Chance-Constraints}


It can be inferred from the individual radial line graphs of Fig. \ref{fig:3_Hosting_Cap_Graphs} that for the hybrid and remote models, transformers supporting loads with PV consistently showed higher EV HC potential (by $\geq$$30\%$). This is because the availability of PV
enabled
coordinated charging during excess PV generation periods. Coordinated charging
also mitigated reverse power flow concerns (see Section \ref{CS2} for more details).

A comparison across all three radial line graphs indicated that the chance-constrained formulation consistently supported more EVs (often by $\geq$$50\%$) than the robust formulation. This implied that with a slight relaxation of the constraints, EV HC potential of the entire feeder can be considerably increased.

Next, we analyze 
the impact of a \textit{mixed} model, which is a combination of the three work schedules described so far, through two case-studies.
The mixed work arrangement was
chosen according to the U.S. Survey of Working Arrangements and Attitudes (SWAA), which states that approximately 60\% of full-time employees work in-person, followed by 27\% in hybrid, and 13\% in remote positions \cite{Measuring_Work_From_Home}.

\begin{figure}[b]
    \vspace{-1em}
    \centering
    \includegraphics[width=0.98\linewidth]{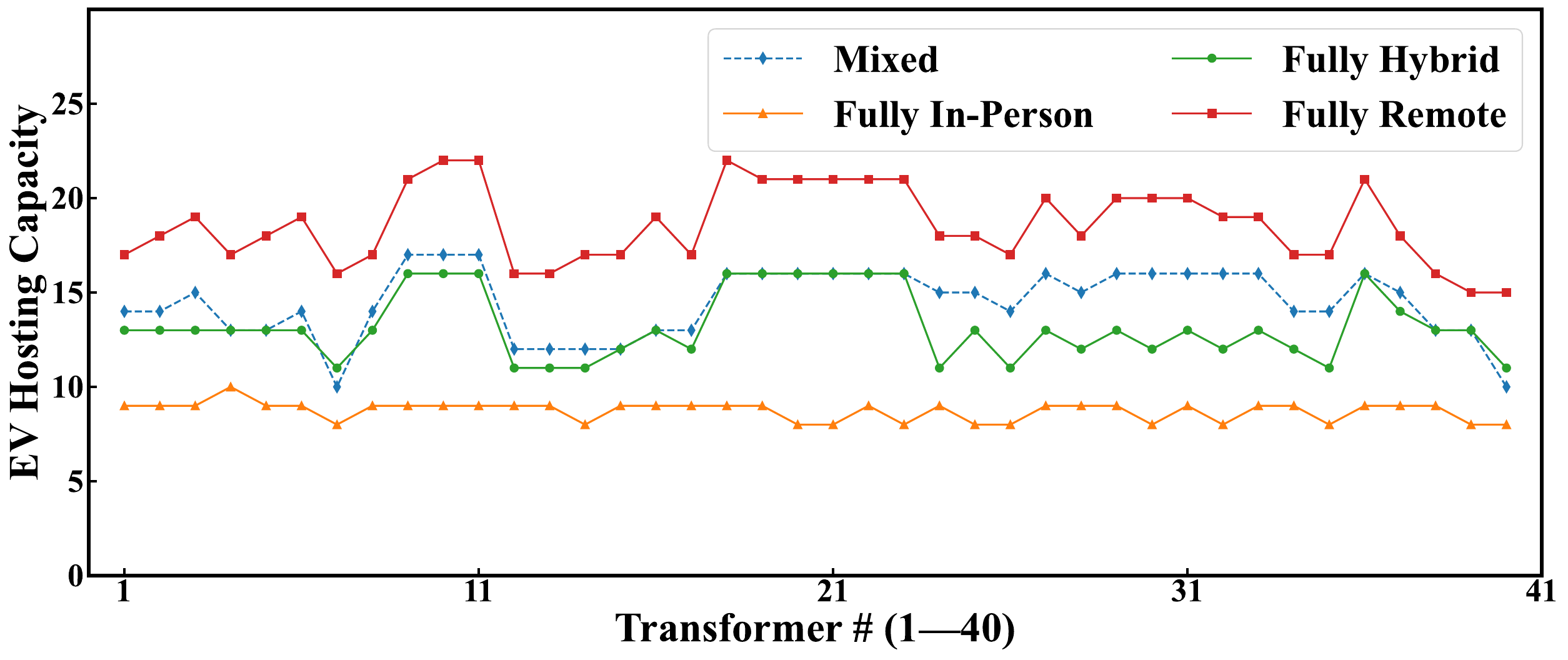}
    \vspace{-1.25em}
    \caption{EV HC results for four
    stochastic 
    optimization models.}
    \label{fig:Hosting-Cap-all-4-Models}
\end{figure}

\begin{figure*}[ht]
    \centering
    \includegraphics[width=0.81\linewidth]{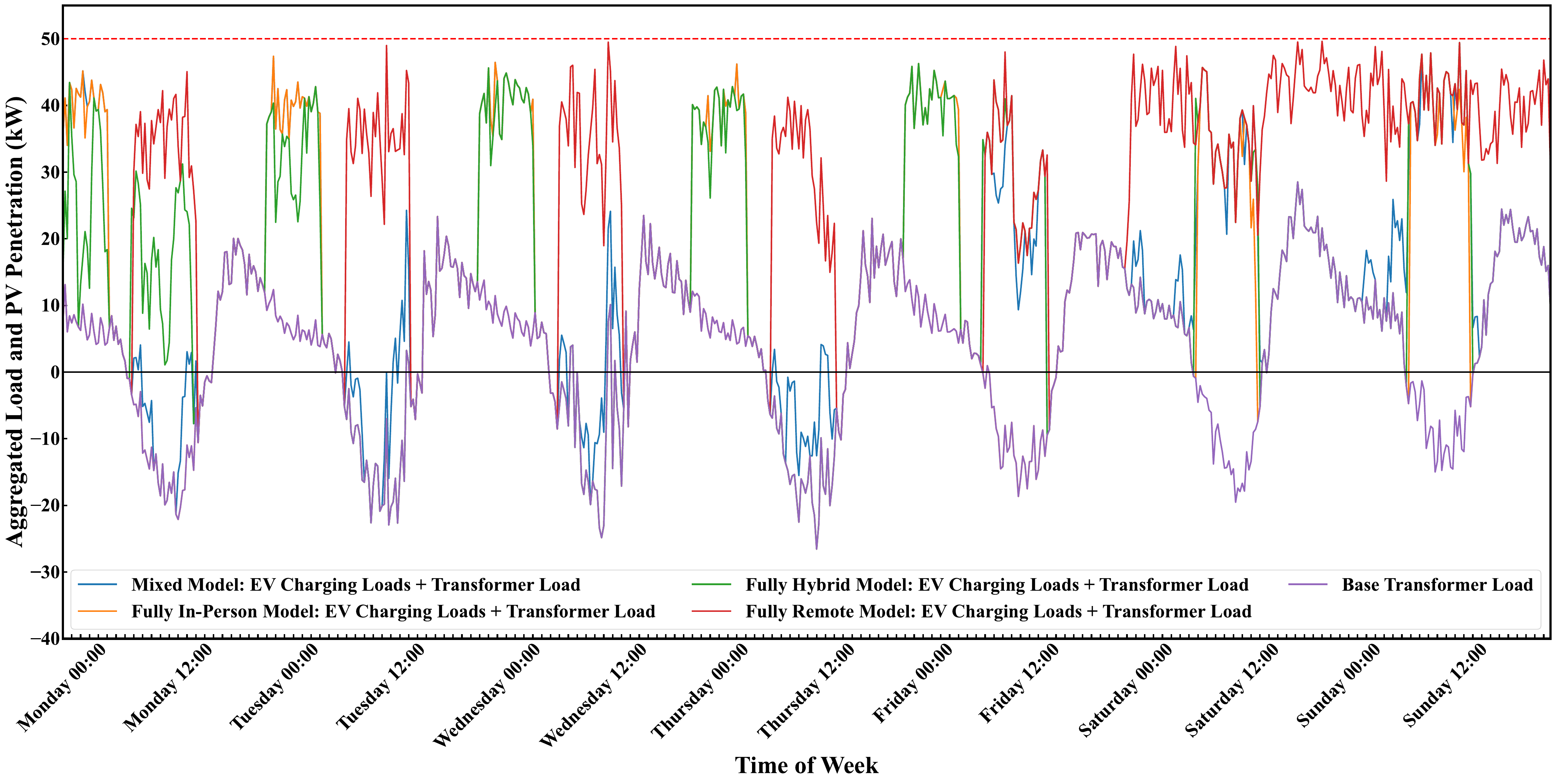}
    \vspace{-1.25em}
    \caption{Charging coordination impact on a transformer supporting loads with PV
    for four EV charging optimization models.}
    \label{fig:Load Impact Figure}
    \vspace{-1em}
\end{figure*}




\vspace{-0.25em}
\subsection{Case-Study 1: Mixed EV Charging Optimization Model}

Fig. \ref{fig:Hosting-Cap-all-4-Models} compares the EV HC results for the chance-constrained formulation for the 40 transformers analyzed in Fig. \ref{fig:3_Hosting_Cap_Graphs}.
From Fig. \ref{fig:Hosting-Cap-all-4-Models}, it can be realized that 
the mixed model has an EV HC range similar to the fully hybrid model. This is expected because both models follow a mostly in-person work arrangement with minor variations: 
the hybrid model allowed all EV owners with only one WFH day (which was Monday/Friday in this study), while the mixed model allowed multiple WFH days to a small percentage of the EV owners.

\vspace{-0.25em}
\subsection{Case-Study 2: Impact of Coordinated Charging
on Transformer Loading and Reverse Power Flows}
\label{CS2}

Fig. \ref{fig:Load Impact Figure} illustrates the ability of the proposed optimization schemes to address the issue of reverse power flow by showing a representative weekly loading profile for a transformer operating at unity power factor and serving customers with rooftop PV.
The figure compares five cases—fully in-person, fully hybrid, fully remote, mixed, and the base case without EVs—under the chance-constrained EV coordination framework. As seen from the purple curve, the base loading profile experiences daily reverse power flows between 8 AM and 3 PM.
Coordinated EV charging provides varying levels of mitigation depending on the work arrangement. The fully in-person model (orange line) alleviates the problem during weekends but struggles with it on weekdays.
The fully hybrid model (green line) almost completely eliminates reverse power flow on the WFH day (Monday/Friday), but shows limited improvement on the remaining weekdays.
The mixed model (blue line) does not fully eliminate reverse power flow but reduces its magnitude across all weekdays.
The fully remote model (red line) mitigates reverse power flows to the maximum extent,
but it also causes the transformer to operate near its capacity at several times during the week.

\section{Conclusion}\label{conclusion}

This paper presents the practical benefits of 
weekly work schedule-aware EV-TOU coordinated charging
on transformer loading.
The analysis of an SRP feeder indicated
that (i) work arrangement flexibility in combination with PV generation boosted EV HC potential by at least 30\%, and (ii) for each work arrangement, the chance-constrained formulation supported approximately 50\% more EVs than its robust counterpart. 
Transformers with lower EV HC ranges were also identified in this study as candidates for upgrades/replacements (e.g., Transformer \#7 in Fig. \ref{fig:3_Hosting_Cap_Graphs}). 
Finally, increased daytime vehicle availability (for charging purposes) was shown to play a decisive role in mitigating PV-driven reverse power flows. 
The  optimization schemes developed in this paper effectively integrate diverse work arrangements, transformer loads, 
cost, and daily commute routines of EV owners, to manage changing grid demands.
The paper also serves as a baseline comparison for future exploration that can be done in this area. 


\bibliographystyle{IEEEtran}
\bibliography{bibref}

\end{document}